%% file: main.tex
\documentclass[11pt, oneside]{article}
\input{preamble.tex}
\usepackage{hyperref}
\usepackage{multirow}
\usepackage{threeparttable}
\usepackage{arydshln}
\usepackage{xcolor}
\usepackage{booktabs}
\usepackage{todonotes}
\usepackage{orcidlink}
\usepackage[most]{tcolorbox}
\usepackage{float}
\makeatletter
\define@key{todonotes}{NA}[]{%
    \setkeys{todonotes}{author=NA,color=green}}%
\makeatother
\usepackage{graphicx} 

\usepackage{caption}
\usepackage{subcaption} 
\usepackage{tikzfigure}

\usepackage{booktabs,adjustbox,pifont}
\newcommand{\cmark}{\ding{51}}  
\title{\textsf{Less can be more for predicting properties with large language models}}
\input{authors} 

\begin{document}

\definecolor{implicationbg}{RGB}{245, 245, 245}
\newtcolorbox{practicalbox}[1][Practical Implications]{%
  enhanced,
  breakable,
  colback=implicationbg,
  colframe=mattext_red,
  fonttitle=\bfseries,
  coltitle=mattext_red,
  attach boxed title to top left={yshift=-\tcboxedtitleheight/2, xshift=1em},
  boxed title style={size=small, colback=white, colframe=mattext_red, sharp corners},
  top=12pt,
  bottom=6pt,
  left=8pt,
  right=8pt,
  title=#1,
}

\begin{refsection}

\maketitle

\begin{abstract}
Predicting properties from coordinate-category data---sets of vectors paired with categorical information---is fundamental to computational science.
In materials science, this challenge manifests as predicting properties like formation energies or elastic moduli from crystal structures comprising atomic positions (vectors) and element types (categorical information). While \glspl{llm} have increasingly been applied to such tasks, with researchers encoding structural data as text, optimal strategies for achieving reliable predictions remain elusive. 
Here, we report fundamental limitations in \gls{llm}'s ability to learn from coordinate information in coordinate-category data. 
Through systematic experiments using synthetic datasets with tunable coordinate and category contributions, combined with a comprehensive benchmarking framework (\mattext) spanning multiple representations and model scales, we find that \glspl{llm} consistently fail to capture coordinate information while excelling at category patterns. 
This geometric blindness persists regardless of model size (up to 70B parameters), dataset scale (up to 2M structures), or text representation strategy. 
Our findings suggest immediate practical implications: for materials property prediction tasks dominated by structural effects, specialized geometric architectures consistently outperform \glspl{llm} by significant margins, as evidenced by a clear \enquote{GNN-LM wall} in performance benchmarks. 
Based on our analysis, we provide concrete guidelines for architecture selection in scientific machine learning, while highlighting the critical importance of understanding model inductive biases when tackling scientific prediction problems.
\end{abstract}

\section{Introduction}

\Glspl{llm} have matched or exceeded human expert performance across diverse scientific domains.\autocite{bommasani2021opportunities} 
This success has prompted researchers to apply \Gls{llm} to property prediction tasks in chemistry and materials science, with several studies reporting encouraging results.\autocite{Mirza_2025, Jablonka_2024, rubungo2023llm0prop0} 
Yet, these achievements are difficult to reconcile with documented failure modes---some as basic as counting letters in a word.\autocite{BerglundTKBSKE24, mirzadeh2024gsm, miret2024llms}

ML models in scientific applications encounter data that contains both coordinate information (vectors) and category information (discrete categorical labels or types), illustrated in \cref{fig:spatial_semantic_asymmetry}. 
\coordinatecategoryfigure
Property prediction from coordinate-category data (data combining vectors with discrete categorical information) is an important example of a scientific prediction task.
It represents a fundamental challenge across the physical sciences.
The task requires predicting properties (such as energies, decay rates, elastic moduli) from data where vector information (e.g., atomic positions, momentum vectors) is paired with category information (e.e., atom types, molecular labels, particle classifications). 
Materials science, for example, has long relied on domain-specific models with hand-crafted features or \gls{gnn} for these predictions\autocite{Reiser_2022, Choudhary_2022}. 
Now, researchers increasingly turn to \glspl{llm} as a general-purpose alternative. 
This shift raises a critical question: Can \gls{llm} effectively learn the coordinate relationships encoded in such coordinate-category data? 
Understanding capabilities and limitations of \gls{llm}'s abilities has profound implications for materials design and discovery, where accurate property prediction accelerates the search for novel materials. 

In this work we show that \modelname{Transformer}-based models, including both causal and masked language models, systematically fail to learn from coordinate information on coordinate-category data. 
Through controlled experiments using synthetic datasets where we can precisely tune the importance of the coordinate information (e.g., atom position vectors) vs. category information (e.g., atom types), we discover that \glspl{llm} are drastically less sample-efficient than conventional approaches when learning from coordinate data.

Our findings reveal a fundamental limitation: while \glspl{llm} excel at learning from the category information (which elements are present), they struggle with coordinate relationships (how atoms are arranged in space). 
This modeling gap persists across model architectures, scales, and training strategies.  Systematically understanding when and why \glspl{llm} succeed or fail at scientific prediction tasks is essential for their reliable deployment.
Our results provide a framework for understanding when to deploy \glspl{llm} versus specialized models for scientific tasks. 
Rather than viewing \glspl{llm} as universal replacements for domain-specific approaches, our work delineates their capabilities and limitations, guiding researchers towards more effective modeling strategies for diverse property prediction tasks.

\section{Results}

\subsection{Probing LLMs with coordinate-category information mixtures}

To understand how \glspl{llm} perform in predicting data containing both coordinate and category information, we created synthetic datasets. 
Each data point consists of 3D coordinates (coordinate information, analogous to position of atoms) and an associated categorical label (analogous to atom types).  
These datasets are inspired by materials or molecular property prediction tasks, where the objective is to predict the property from coordinates and atom types.

We computed scalar target values for these coordinate-category data points using a physics-inspired hypothetical potential function (which can be thought of as an energy function). 
To that end, we construct the hypothetical potential such that we can continuously tune the impact of categories (i.e., types) vs.\ coordinate information, see \cref{eq:hypo_pot}. We give a more detailed explanation of \cref{eq:hypo_pot} in \cref{subsec:hypothetical_potential}.

\begin{equation}
        E =\alpha E_\mathrm{category} +(1-\alpha) E_\mathrm{coordinate} \quad \quad \alpha \in (0,1)
    \label{eq:hypo_pot} 
\end{equation}

In the extreme of $\alpha=0$ the labels depend solely on $E_\mathrm{coordinate}$ (the energy contribution from the coordinate information, i.e.\ the positions of the points in space), while for $\alpha=1$ the labels depend only on $E_\mathrm{category}$ (the energy contribution from the category information, i.e. the types of points). 
Based on \Cref{eq:hypo_pot} and our goal of elucidating effects from coordinate and category data types, we generate synthetic datasets with diverse coordinate and category configurations. 
Some datasets have points arranged in regular patterns (for e.g., cube corners), others have other coordinate configurations, and also datasets vary in their number of category types. 
For each dataset, we generate multiple versions by varying $\alpha$ (e.g. $\alpha$ = 0, 0.2, $\cdots$, 1) that span the spectrum from purely coordinate-dependent to purely category-dependent energy landscapes.
This controlled $\alpha$-sweep framework allows us to isolate and study how \glspl{llm} learn from coordinate versus category information, while ensuring our findings generalize across various coordinate geometries and category distributions.

\begin{figure}[htb]
    \centering
    \includegraphics[width=14cm]{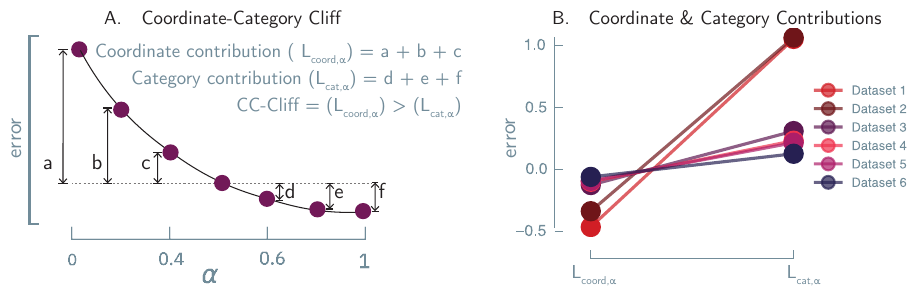}
    \caption{\textbf{Illustration of Coordinate-Category Cliff (\gls{ccc}) (A). Coordinate contribution (\gls{coc}) and Category contribution (\gls{cac} for different datasets (B)}. The coordinate contribution and category contribution are defined as: 
    $L_{\mathrm{coord}, \alpha}= \sum_{\alpha \in \alpha_g} \text{loss}(\alpha) - 3 \times \text{loss}(0.5)$ and $L_{\mathrm{cat}, \alpha} = \sum_{\alpha \in \alpha_c} \text{loss}(\alpha) - 3 \times \text{loss}(0.5)$, where $\alpha_g = \{0, 0.2, 0.4\}$ and $\alpha_c = \{0.6, 0.8, 1.0\}$. Accounting for the three $\alpha$ values included in each sum we subtract $\text{loss}(0.5)$ three times (which effectively means we subtract the loss at $\alpha$=0.5 from each individual loss contributing to the sum).  The plot on the right shows the \gls{coc} and \gls{cac} for six different datasets. For all datasets, the language model shows a positive \gls{ccc}, i.e., \gls{coc} $>$ \gls{cac}. Which is a gap in learning coordinate computations compared to category computations. There is also variation in the cliff magnitude across different datasets.
    }
    \label{fig:ssa}
\end{figure}

Using these datasets, we follow established property prediction procedures \autocite{GruverSMWZU24,Jablonka_2024} and finetune \modelname{BERT} \autocite{devlin2018bert} models to predict the labels for all versions of the dataset (each with a different $\alpha$ value). To quantify how well the models learn from each type of information, we compute two aggregate error metrics: the \gls{coc} and the \gls{cac}. As illustrated in \cref{fig:ssa}A, \gls{coc} measures the model's error on tasks dominated by coordinate information (i.e., for $\alpha$ values close to 0), while \gls{cac} measures the error on tasks dominated by category information (i.e., for $\alpha$ values close to 1). More details on are present in \cref{subsec:ssa}. If \gls{coc} $>$ \gls{cac}, we call this phenomenon the \gls{ccc}---the systematic underperformance on tasks requiring coordinate information. 

Our key finding, shown in \cref{fig:ssa}B, is that \gls{coc} is consistently and significantly larger than \gls{cac} across all datasets. We find that models consistently perform much worse in learning from coordinate information than from category information. This is consistent across different datasets (which are samples from different distributions of coordinate-category configurations). For the purpose of quantifying this metric, we can compute \gls{ccc} as \gls{coc} $-$ \gls{cac}.
A \gls{ccc} of zero would indicate equal performance, while a positive gap reveals a bias against learning from coordinate data. These findings indicate that language models might not be the best choice for all problem settings.

\subsection{Comparison with \modelname{n-gram} models}

To better understand the source of those problems, we performed the same experiments with \modelname{n-gram} models. 
This experiment is based on the fact that it has been shown that some behaviors of \glspl{llm}, such as how it uses its context, can be explained by thinking of them as \modelname{n-gram} models.\autocite{nguyen2024understanding} 

In addition, we change the complexity of the learning task\autocite{zhou2023algorithms} by binning the pairwise distance in coordinate contribution to energy ( $E_\mathrm{coordinate}$)--- modeling the task as an increasingly complex multiclass classification problem (with the large number of bins approaching the regression problem). 
More detailed explanation on the rationale for this experiment and methodological details is in \cref{subsubsec:ngram_rationale} and \cref{subsubsec:ngram_task_complexity}, respectively.
\begin{figure}[htb]
    \centering
    \includegraphics[width=1\linewidth]{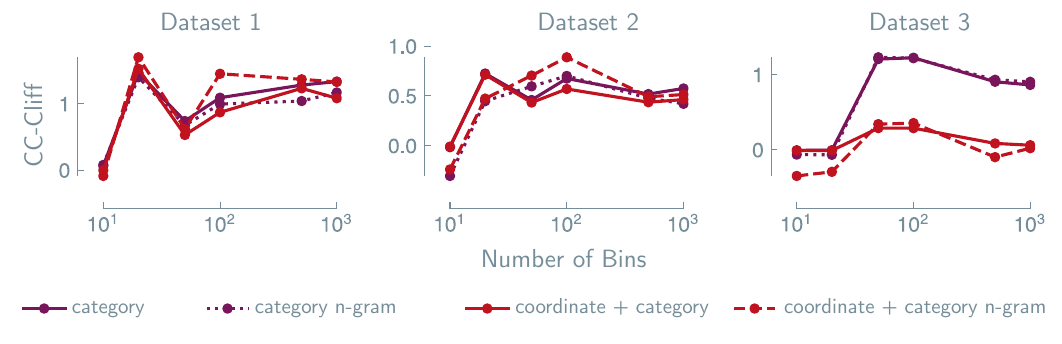}
    \caption{ \textbf{\gls{ccc} in hypothetical potential energy prediction as a function of binning the potential across different datasets. }
    The figure illustrates the \gls{ccc} for models tasked with predicting hypothetical potential energy. The analysis is performed across three distinct datasets. The \gls{ccc} is plotted against the number of bins used to discretize the pairwise distance (logarithmic scale). Fewer bins correspond to a coarser, generally easier potential energy prediction task for the structures within that dataset, while a higher number of bins represents a finer-grained, more challenging prediction task. Each line corresponds to a different text input (with different levels of information) used to describe the data points in the dataset: categorical (purple), and coordinate and categorical (red). Solid lines indicate \modelname{Transformer}-based models, and dotted lines represent their \modelname{n-gram} counterparts. The plot shows almost identical behavior of language models to that of \modelname{n-gram} models suggesting that \glspl{llm} inherit the properties of \modelname{n-gram} for tasks involving coordinate and category data.
    }
    \label{fig:ngram}
\end{figure}

In \Cref{fig:ngram} we find that for all datasets we analyzed, the behavior of \modelname{n-gram} models closely follows that of the \glspl{llm} trained on only category or full category and coordinate information.
This showcases that for this task the behavior we observe can be approximated with \modelname{n-gram}s. 
This understanding is very instructive because for \modelname{n-gram} models we do not expect to be able to predict properties that dependent on positional information as they do not account for this type of data. 
We would rather expect that we would need combinatorially many samples to describe such data with \modelname{n-gram} models. 
Our findings from the attention analysis further reveals that coordinate information is not leveraged by the models while making prediction (more details in Appendix \cref{appendix_sec:attention_analysis}). 
Taken together, this provides compelling evidence for \gls{llm}'s failure to properly account for coordinate information.

\subsection{Implications for materials property predictions}

\begin{figure}[H]
    \centering
    \includegraphics[width=1\linewidth]{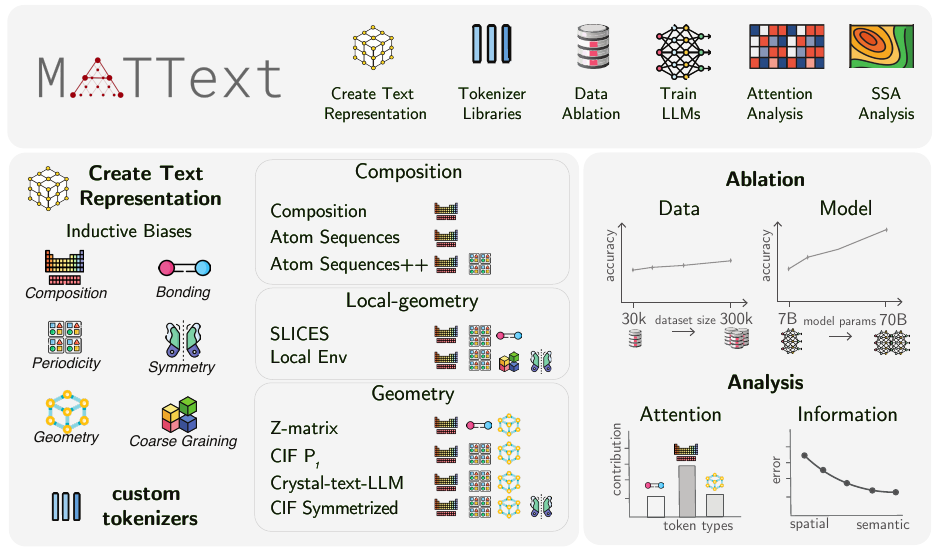}
    \caption{\textbf{The above figure outlines an overview of the MatText framework.} MatText is a holistic platform that supports end-to-end language modeling of materials, creation of representations, model training, and streamlined analysis of results. MatText enables the creation of text representations of crystal structures, offering nine different representations, each with distinct inductive biases. These inductive biases explicitly encode diverse types of information, such as bonding, periodicity, symmetry, and other shown in the middle section. The framework also support various tokenization methods, such as atom-level and representation-specific tokenization, as well as different ways to tokenize numbers. MatText facilitates pretraining and finetuning of both causal and masked language models, and features modules for scaling up model and data sizes. Additionally, MatText provides tools for analyzing attention mechanisms, assessing the contribution of tokens in predictions based on attention scores, and performing \gls{ccc} analysis using hypothetical potentials.}
    \label{fig:mattext-overview}
\end{figure}

Those finding have concrete practical implications for materials property prediction and various other machine learning tasks.  
To investigate these implications, we built an open source framework, \mattext, with which we can represent materials in various text form and seamlessly test the performance of different language models on different materials tasks (see \Cref{fig:mattext-overview}). 
To avoid that our results are confounded by a special representation, model architecture, tokenizer, or dataset choice, we designed targeted experiments to understand the effects of all of those elements.

\paragraph{Representation} One of the most important factors in determining the performance of models in materials science is the data representation. 
The development of text-based representations for materials is an active area of research, and there exists no universally accepted \enquote{best} text-based representation for materials (more details on representations are in \cref{subsec:material_modelling}). 
With the aim of providing a broad collection in MatText, we implemented many previously proposed text representations, and created several novel ones to probe specific a comprehensive set of inductive biases.
Thus, the representations in MatText not only differ in the way they present the information in text, but also in the information they contain (see \Cref{fig:mattext-overview}). 
Some, such as \slices\autocite{xiao2023invertible} only contains information about the composition and bonds. 
Others, such as \cifsym contain information about the position of all atoms, in addition to unit cell and symmetry information. 
A full description of all representations we implemented can be found in Appendix \cref{appendix_sec:representation}

\begin{figure}[H]
    \centering
     \includegraphics[width=\textwidth]{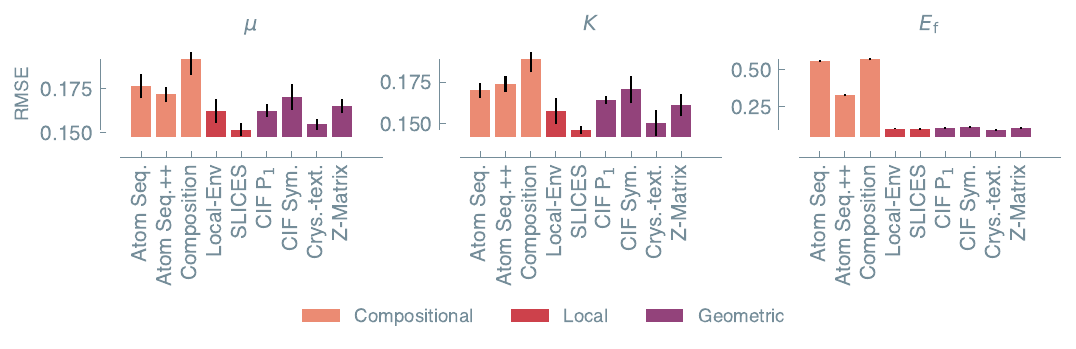}
    \caption{\textbf{Predictive performance of \glspl{llm} in materials property prediction tasks (shear modulus ($\mu$), bulk modulus ($K$), and perovskite formation energy ($E_f$)) for different representations.} Performance is measured by \gls{rmse}, with lower values indicating better performance. Representations are grouped by type: compositional-based (orange), local environment-based (red), and geometry-based (purple). Across all three properties, local environment based representations generally achieve the best performance, while explicit geometric representations show limited improvement or even degraded performance. Notably, the \slices representation, which lacks explicit coordinate information, performs comparably to geometry-aware representations like \cifpone, suggesting that current \glspl{llm} do not effectively leverage explicit coordinate information for materials property prediction. The error bars indicate the standard deviation across five-fold cross-validation. A notable exception is the perovskites dataset, where there is a big difference between representation.  This dataset has few unique chemical environments compared to the shear and bulk modulus datasets (see \Cref{fig:structural_variations_all_property}), wherefore most of the variance cannot be explained using composition information alone. }
    \label{fig:representation-depedence}
\end{figure}

In \Cref{fig:representation-depedence} we show the predictive performance on materials property prediction tasks as a function of the input representation. 
We can make an interesting observation: In many cases, the addition of information does not improve the model performance. 
For instance, we observe \slices performing better than \cifpone for all the properties, also \localenv shows comparable performance in prediction of bulk modulus and shear modulus.
These results support our observation that coordinate information cannot be effectively used by \glspl{llm}. 
Appendix \cref{appendix_sec:tokenizer} shows that these findings are not specific to the choice of tokenizers or how we tokenize numbers.

\paragraph{Scale}

\begin{figure}[H]
    \centering
    \includegraphics[width=\textwidth]{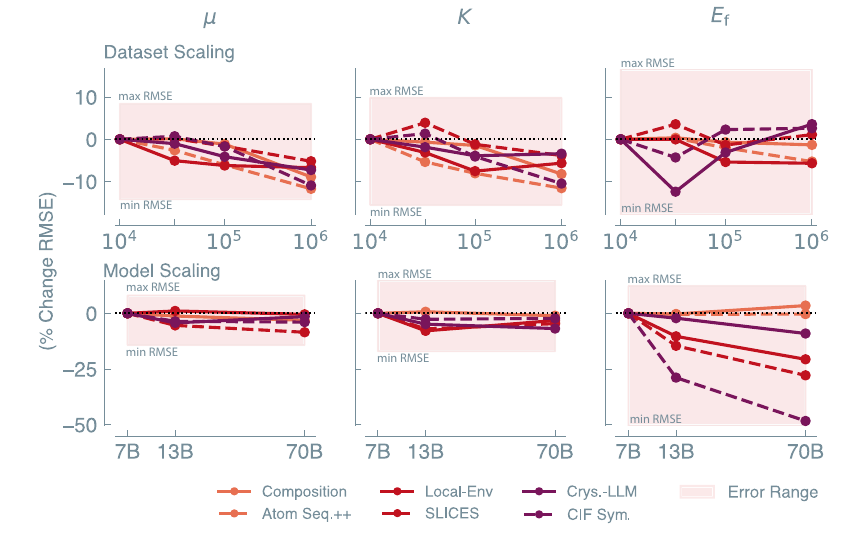}
    \caption{ \textbf{Impact of dataset and model scaling on predictive performance of material representations. }
    This figure illustrates the percentage change in \gls{rmse} relative to a baseline (on the left side), for various material representations when scaling (top row) the pretraining dataset sizes for \modelname{BERT}\autocite{devlin2018bert} models and (bottom row) the language model parameter count (7B, 13B, 70B parameters) for instruction-tuned LLaMA\autocite{touvron2023llama} model. Each column corresponds to a different material properties shear modulus ($\mu$), bulk modulus ($K$), and formation energy ($E_f$). Each colored line tracks the mean \gls{rmse} percentage change for a specific material representation as either dataset size or model size increases. The baseline for dataset scaling is the mean \gls{rmse} achieved with 30,000 structures dataset for that representation. For model scaling, the baseline is the mean \gls{rmse} of the 7 billion (7B) parameter model for that representation. Negative percentage changes indicate improved predictive performance (lower \gls{rmse}) compared to the respective baseline. The light red shaded region in each subplot defines the overall range of observed \gls{rmse} percentage changes across five folds. The upper and lower boundaries represent the global maximum and minimum percentage changes in \gls{rmse}, respectively, from all individual cross-validation fold results of all representations. This band thereby encapsulates the full spectrum of performance variation encountered for that property under the given scaling regime. }
    \label{fig:scale}
\end{figure}

One of the most widely established ways of improving model performance is by scaling up --- i.e. to train larger models on more data\autocite{kaplan2020scaling,hoffmann2022training} --- which has been proposed in general language modeling as well as scientific domains including chemistry \autocite{frey2023neural}.  
While there have been some inconsistencies \autocite{mckenzie2023inverse}, the general observation on many tasks has been that scale reliably improves performance --- a pattern that has been formalized with empirical scaling laws \autocite{hoffmann2022training}.
To analyze if the performance in predicting materials properties can be improved by scaling, we ablated  dataset size and model parameters count. 
We evaluated all our experiments, including those with larger model and data scales, in a five-fold cross validation setting to measure variance. 
To our knowledge, this is the first study to conduct such a comprehensive and extensive benchmarking analysis leveraging more than 2000 language model training jobs.
\Cref{fig:scale} showcases our findings: We show the impact of dataset scaling in the top row and model size scaling in the bottom row. 
In all plots, we measure the change in \gls{rmse} with respect to the starting configuration. 
We indicate the range between the biggest and smallest change in \gls{rmse} with a colored rectangle---any point within this range is not distinguishable from a random fluctuation within our experimental setting.

Remarkably, we find that for all representations and prediction targets we investigated, there is no effect of either scaling dataset size or model parameter count (we provide more extensive details in \cref{subsec:material_modelling}).  
Our findings indicate that adding scale in data or model size is not enough to sizably improve the modeling of material properties using \glspl{llm}.

\begin{figure}[H]
    \centering
    \includegraphics[]{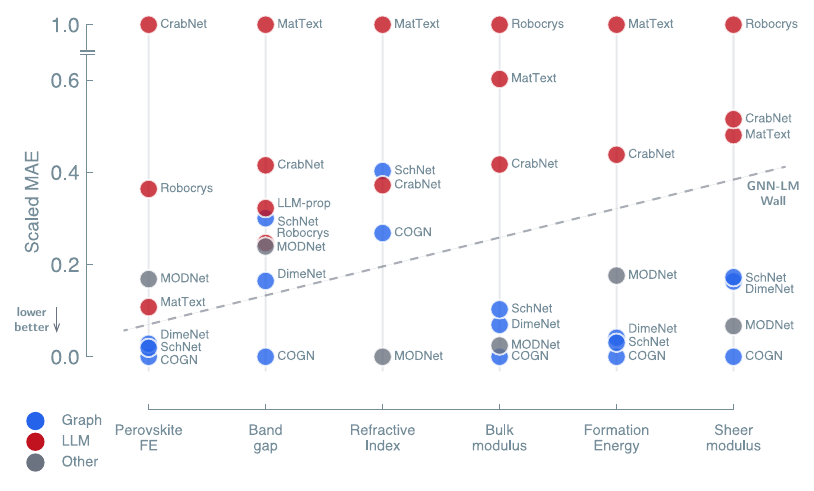}
    \caption{\textbf{The \gls{gnn}-LM wall} The figure presents a comparative performance analysis, measured by scaled \gls{mae}, between graph-based models (red circles) and language-based models (blue circles) across six different material property prediction tasks. The dashed grey line, labeled \enquote{\gls{gnn}-LM wall,} serves as a visual guide to generally demarcate the performance difference between the two classes of models. \gls{gnn} models (shown in blue) consistently achieve better performance across all properties, while \gls{llm} models (shown in red) generally exhibit higher errors. This systematic performance gap suggests that current \gls{gnn} architectures are more effective at learning and predicting materials properties compared to \gls{llm}-based approaches. Shown in blue are \glspl{gnn} based approaches (\modelname{COGN}\autocite{ruff2024connectivity}, \modelname{SchNet}\autocite{schtt2017schnet}, \modelname{DimeNet}\autocite{johannes2020di}) and in red are approaches based on language models (\modelname{LLM-prop}\autocite{rubungo2023llmprop}, \modelname{CrabNet}\autocite{wang2021compositionally}, \modelname{Robocrys} \autocite{qu2024leveraging}). Additionally, \modelname{MODNet}\autocite{de2021robust} is included, which is an approach distinct from both \glspl{gnn} and \glspl{llm}}
    \label{fig:llm-gnn-wall}
\end{figure}

\paragraph{The \gls{gnn}-LM wall}

Given our findings and the recent surge of attempts of using \glspl{llm} for (material) property predictions, we analyzed the commonly used MatBench leaderboard \autocite{Dunn_2020} to further understand relevant consistencies and differences. 
In \cref{fig:llm-gnn-wall}, we show the MAE relative to the worst performing among the selected approaches here for various approaches. 
We distinguish with colors different modeling paradigms: Showing in blue approaches relying on \glspl{gnn} and in red approaches based on language models. 

Notably, we consistently find language-modeling based approaches at the top of the plot, indicating the highest error and lowest performance. 
Additionally, we find a striking segregation, where one can almost draw a \enquote{line} separating \gls{gnn}-based approaches from language-modeling based ones.
This observation underscores our findings that certain properties --- especially if they depend on coordinate information (e.g., mechanical properties) --- cannot efficiently be modeled with current language modeling approaches: Independent of representation, model size, and dataset size.  
In practice, more fundamental changes to the architecture of language models are needed to make it possible to model coordinate information with greater effectiveness. 
Alternatively, researchers can already rely on model architectures that are optimized to deal with such data.

\section{Conclusions}

Predicting properties from coordinate-category data---whether atomic positions in materials, molecular coordinates in chemistry, or feature vectors in machine learning---represents a fundamental challenge across the physical sciences. 
The promise of \glspl{llm} has captivated researchers with their apparent universality: a single architecture that learns from text alone, yet seemingly grasps complex patterns across domains.
This vision has driven a surge of attempts to deploy \glspl{llm} for scientific property prediction, with researchers racing to encode crystal structures, molecules, and other vectorial data into text, hoping to unlock the same transformative capabilities witnessed in natural language processing.

Our systematic investigation reveals a more sobering reality. 
Through controlled experiments with synthetic datasets where we can precisely tune the balance between category and coordinate information, we uncover a fundamental limitation: \Glspl{llm} systematically fail to learn from coordinate information in coordinate-category data. 
This failure persists across architectures, scales, and training strategies. Even massive increases in data and model parameters---the traditional levers of improvement in language modeling---yield no meaningful gains. 
\glspl{llm} behave remarkably like \modelname{n-gram} statistics, capturing categorical patterns while remaining blind to the spatial arrangements that often determine physical properties.

Yet this apparent failure point to a path forward. 
By delineating where \glspl{llm} excel (learning from categorical information like element types) and where they struggle (extracting geometric relationships), we provide a framework for making informed modeling choices. 
Our findings suggest that rather than pursuing \glspl{llm} as universal replacements for specialized architectures, the field should embrace a more nuanced approach: deploying \glspl{llm} for tasks dominated by compositional patterns while reserving geometric neural networks and other purpose-built architectures for problems requiring spatial reasoning.

In an era rushing toward universal models, we demonstrate that understanding the inductive biases of our tools remains as crucial as ever. 
The future of scientific machine learning might therefore not lie in finding one universal model to rule them all, but in matching the right architecture to the right problem.

\section{Methods}

\subsection{Modeling coordinate information with \modelname{Transformers}}


Benchmarking with real-world systems can be confounded by inherent correlations within the training data. 
For example, certain compositions might predominantly form specific structural motifs or crystallize in particular space groups in materials. 
A model could leverage these statistical correlations to predict a structure-dependent property accurately, even with a limited understanding of the actual geometric arrangements, by primarily relying on compositional cues. To illustrate, if a specific elemental combination almost exclusively forms a known ground state structure, a model might appear to \enquote{understand} the geometry leading to this ground state by simply recognizing the composition. 
This makes it challenging to isolate and quantify a model's proficiency in processing coordinate information. 
To address this, we designed a hypothetical potentials, where the \enquote{ground truth} labels can be continuously tuned to depend more or less on categorical versus positional features. 
This allows for a systematic probing of a model's learning capabilities across this spectrum.

\subsubsection{The Physics inspired-hypothetical Potential}
\label{subsec:hypothetical_potential}

We consider a system characterized by a collection of $N$ entities, where each entity $k$ possesses a category type (e.g., discrete type) and a continuous coordinate attribute $\mathbf{r}_k$ (e.g., position). 
We then define a scalar label, $E$, analogous to an energy, for each system configuration. 
This label is a linear interpolation between a purely category term ($E_\mathrm{category}$) and a purely coordinate term ($E_\mathrm{coordinate}$), controlled by a mixing parameter $\alpha \in [0,1]$:

\begin{equation}
    E(\alpha) = \alpha E_\mathrm{category} + (1-\alpha) E_\mathrm{coordinate}
    \label{eq:hypo}
\end{equation}

The contribution from category related information, $E_\mathrm{category}$, is defined as a sum over the counts of each entity type:
\begin{equation}
    E_\mathrm{category} = \sum_{k=1}^{N_t} w_k n_k
    \label{eq:hypo_semantic}
\end{equation}

where $N_t$ is the number of unique entity types, $n_k$ is the count of entities of type $k$, and $w_k$ is a weight associated with particles of type $k$ (in our calculations we maps chemical elements to their energy parameters from the \gls{uff}\autocite{rappe1992uff} ). 
This term represents the contribution of the intrinsic properties of each type of particle. $E_\mathrm{category}$ term captures contributions based solely on the \enquote{what} and \enquote{how many} of the entities.

The contribution of coordinate-related information, $E_\mathrm{coordinate}$, is designed to depend on the vectors and their relative arrangements:
\begin{equation}
    E_\mathrm{coordinate} = \sum_{i=1}^N \sum_{j \in \mathcal{N}(i)} V\left(\left|\mathbf{r}_i-\mathbf{r}_j\right|\right)
    \label{eq:hypo_spatial}
\end{equation}

where  $V(|\mathbf{r}_i-\mathbf{r}_j|)$ represents a pairwise interaction potential (e.g., Lennard-Jones in our experiments) dependent on the distance between entity $i$ and its neighbors $j$ within a defined neighborhood $\mathcal{N}(i)$. 
This term captures contributions based on the \enquote{where} and \enquote{how arranged} of the entities.

By varying the mixing parameter $\alpha \in [0,1]$, we can continuously tune the nature of the generated labels. 
In one extreme, at $\alpha=0$, the label $E$ becomes solely dependent on coordinate information ($E = E_{\mathrm{coordinate}}$). Conversely, at the other extreme, $\alpha=1$, the label depends exclusively on category information ($E = E_{\mathrm{category}}$). Intermediate values of $\alpha$, such as $\alpha=0.5$, yield labels that represent a balanced contribution from both positional and category terms, while other values within the $(0,1)$ interval allow for a spectrum of relative influences between these two components. 
This setup, while a simplification of complex real-world interactions, provides a controlled environment to assess how well a model learns from these distinct information modalities.

\subsubsection{Coordinate-Category Cliff}
\label{subsec:ssa}

To quantify the model's relative ability to learn from coordinate versus category information, we define two aggregate metrics: the \textbf{coordinate contribution (\gls{coc})} and the \textbf{category contribution (\gls{cac})}. 
These scores quantify the model's performance drop in regimes dominated by either coordinate or category information, relative to its performance in a balanced scenario ($\alpha=0.5$).

The coordinate contribution (\gls{coc}) is calculated by summing the losses for $\alpha$ values where coordinate information dominates, and subtracting a baseline derived from the loss at $\alpha=0.5$:
\begin{equation}
    L_{\mathrm{coord}, \alpha} = \left( \sum_{\alpha \in \alpha_g} \text{loss}(\alpha) \right) - |\alpha_g| \cdot \text{loss}(0.5)
    \label{eq:spatial_contribution}
\end{equation}
where $\alpha_g = \{0, 0.2, 0.4\}$ is the set of $\alpha$ values biased towards coordinate information, and $|\alpha_g|=3$ is the number of points in this set.

Similarly, the category contribution (\gls{cac}) is calculated for $\alpha$ values where category information dominates:
\begin{equation}
    L_{\mathrm{cat}, \alpha} = \left( \sum_{\alpha \in \alpha_c} \text{loss}(\alpha) \right) - |\alpha_c| \cdot \text{loss}(0.5)
    \label{eq:semantic_contribution}
\end{equation}
where $\alpha_c = \{0.6, 0.8, 1.0\}$ is the set of $\alpha$ values biased towards category information, and $|\alpha_c|=3$.

A higher contribution (e.g., a large positive \gls{coc}) indicates that the model struggles more in that respective regime (coordinate-dominant for \gls{coc}) compared to the balanced $\alpha=0.5$ case. 
Conversely, a score near zero suggests performance comparable to the balanced case for that regime.
The \gls{ccc} is then analyzed by comparing \gls{coc} and \gls{cac}. For instance, in  \cref{fig:ssa}, \gls{coc} and \gls{cac} is plotted against each other.

\begin{itemize}
    \item If \gls{coc}$\gg$ \gls{cac}, the model finds it substantially harder to learn from coordinate information than from category information, relative to the $\alpha=0.5$ baseline. If \gls{cac} $ \gg$ \gls{coc}, the converse is true.
    \item If \gls{coc} $\approx$ \gls{cac}, the model exhibits a similar level of difficulty (or ease) in both regimes relative to the baseline.
\end{itemize}
This quantitative framework, based on systematically varying the nature of the learning task, allows for a more nuanced assessment of an \gls{llm}'s capabilities beyond performance on a single, fixed task. 
It specifically probes the model's ability for processing and integrating different fundamental types of information present in vector-based descriptions of systems.

\subsection{Rationale for \modelname{n-gram} comparison and task complexity variation}
\label{subsubsec:ngram_rationale}

Given that our task involves predicting labels based on textual inputs, comparing \gls{llm} performance to that of \modelname{n-gram} models provides an informative baseline.\autocite{nguyen2024understanding} 
Intuitively, \modelname{n-gram} models are not expected to capture complex, non-local geometric dependencies or perform intricate calculations (like those in \cref{eq:hypo_spatial}) without an exponentially large number of samples to cover the combinatorial space of token sequences representing these structures. 
For instance, to accurately model pairwise interactions using only local \modelname{n-gram}s, 
 a purely compositional vocabulary might contain around 128 tokens (for elements and digits). However, to also encode atomic positions with a modest \SI{0.1}{\angstrom} resolution across a \SI{20}{\angstrom} cell, each atom's position adds 8 million ($8 \times 10^6$) possible location identifiers. This inflates the vocabulary of \enquote{atom-at-a-position} tokens to over a billion ($>10^9$).  Learning a simple pairwise interaction (a 2-gram) would require sampling from a space of possibilities that is $(8 \times 10^6)^2 \approx 6.4 \times 10^{13}$ times larger than in the compositional-only case. Learning physical laws this way is thus exceptionally sample-inefficient, as it would require an astronomical amount of data to observe a meaningful fraction of relevant atomic pairings.
 
If \glspl{llm} show similar performance trends to \modelname{n-gram} models on this task, it might suggest they are relying on more superficial textual patterns rather than developing a deeper, generalizable understanding of the underlying relationships.

Concurrently, we investigate how the models' learning capabilities are affected by the \enquote{resolution of the geometric information}. 
The calculation of labels, even in our simplified potential (\Cref{eq:hypo}), involves operations on continuous vector attributes, such as the pairwise distances $|\mathbf{r}_i-\mathbf{r}_j|$ used in $E_{\mathrm{coordinate}}$. 
These operations might not align perfectly with a \modelname{Transformer}'s natural computational style (cf. RASP conjecture \autocite{zhou2023algorithms}), potentially leading the model to learn shortcuts or struggle with precise generalization, especially when the input data (e.g., textual representation of crystal structures) requires complex parsing algorithms to extract these from representations of varying length.

\subsubsection{Task Complexity Variation via Pairwise Distance Binning}
\label{subsubsec:ngram_task_complexity}

To probe this, we modify the generation of the target labels $E(\alpha)$ by discretizing the pairwise distances $|\mathbf{r}_i-\mathbf{r}_j|$ into a varying number of bins ($M$) before they are used to calculate $E_{\mathrm{coordinate}}$ (\cref{eq:hypo_spatial}). 
This effectively changes the \enquote{physical} model whose properties the \gls{llm} must learn:
\begin{itemize}
    \item When $M$ is small (coarse distance binning), the $E_{\mathrm{coordinate}}$ term, and consequently $E(\alpha)$, is derived from a simplified geometric landscape where only gross differences in inter-entity distances matter. The learning task involves predicting labels from a system with inherently lower geometric resolution.
    \item When $M$ is large (fine distance binning), $E_{\mathrm{coordinate}}$ and $E(\alpha)$ are derived from a system that is sensitive to subtle variations in distances, more closely approximating a potential based on continuous geometric inputs.
\end{itemize}
This allows us to assess whether the models are capable of learning fine-grained geometric dependencies when the target labels are sensitive to them, or if their performance degrades, potentially indicating an inability to extract or utilize high-resolution continuous information from the textual input, or an over-reliance on coarser patterns. 
This approach helps to distinguish between a model's ability to learn a complex function versus its ability to learn a function defined by high-resolution input features.

\subsubsection{\modelname{n-gram} Model Implementation}
\label{subsubsec:gram_implementation}

For the \modelname{n-gram} model experiments, the input system descriptions (e.g., CIF files) similar to that used for the \glspl{llm}) were used to generate features. 
Specifically, we extracted \gls{tfidf}\autocite{sparck2004statistical} weighted features from unigrams and bigrams derived from this textual data. 
An \modelname{XGBoost}\autocite{chen2016xgboost} classifier was then trained on these TF-IDF features. The target for these \modelname{n-gram} models was the binned energy label, consistent with the task given to the \glspl{llm} in the binned experiments described below.
The model performance was evaluated using standard classification metrics appropriate for the multi-class problem.

\subsection{\mattext python package}
There has been no tool to systematically derive text representations for materials.
\mattext provides an object-oriented way to convert crystal structures into text representations. 
Each of the text representations has different information content (going from just composition information to information about the composition and the position of all the atoms,  see \cref{fig:mattext-overview}), allowing us to analyze what information language models can use for material property predictions. 
In addition, the representations feature different combinations of inductive biases, thereby enabling us to identify the most meaningful ones in our analysis. 
For analysis purposes in this article we have categorized representations into broadly three categories: 
composition based representations (\comprepgroup), representations that focus on local geometry (\localrepgroup) and finally representations that capture 3D geometry (\geomrepgroup). 
More details about inductive biases and various representation with examples is provided in Appendix \cref{appendix_sec:inductive_biases} and \cref{appendix_sec:representation} respectively.

\subsection{Scaling experiments}

To systematically investigate the impact of scale on performance, we conducted experiments along two primary axes: dataset size and model parameter count. First, we explored data scaling by pre-training \modelname{BERT} models from scratch on datasets of progressively larger sizes. Second, we examined model scaling by applying parameter-efficient fine-tuning to pre-existing \modelname{Llama-2} models of different scales. The specific setups for these two experimental tracks are detailed below.

\paragraph{Scaling dataset sizes} 
\label{subsubsec:scaleup_data}
We created pre-training datasets of four different sizes for all representations shown in \Cref{fig:mattext-overview}. 
Our datasets consist of 30K structures, 100K structures, 300K structures and 2000K structures, i.e. four datasets for all nine representations leading to 36 pre-training datasets all together. We pretrained a separate base \modelname{BERT} model for all these datasets, yielding 36 base \modelname{BERT} models, which are then finetuned on different properties from matminer\autocite{ward2018matminer} (particularly on formation energy of perovskite structures\autocite{castelli2012new}, bulk modulus\autocite{de2015charting} and shear Modulus\autocite{de2015charting}, enabling us to compare with state-of-the-art models in MatBench).

\paragraph{Scaling model size}
\label{subsubsec:scaleup_data}
To assess the influence of model scale, we performed parameter-efficient fine-tuning of \glspl{llm} of different sizes, specifically \modelname{Llama-2 7B}, \modelname{Llama-2 13B} and \modelname{Llama-2 70B} models.\autocite{touvron2023llama}. We finetune the \gls{llm} models with \gls{lora} \autocite{hu2021lora}.
Packing was disabled as we only masked the completion during training, and a data collator was defined to train only for the generation part of the prompts.
In addition to \gls{lora}, we use 4-bit quantization with \texttt{nf4} quantization type and \texttt{float16} compute data type.\autocite{dettmers2024qlora}

\subsubsection{Hyperparamters}

\paragraph{Pre-training} 
We choose a batch size and context length specific (given in \cref{{table:context_lengths}}) to representation and training models for a total of 50 training epochs and a learning rate of \num{2e-4} using a masked language modeling (MLM) approach with a probability of 0.15 for masking tokens. 

\paragraph{Finetuning}

For fine-tuning, we employ early stopping with a patience of 10 and a threshold of 0.001 to prevent overfitting, utilizing 20\% of the data for evaluation while the remaining 80\% is used for training. 
The learning rate is set to \num{2e-4}. 
The pretrained base model layers are not frozen, and a regressor head on top of the base model is used for the regression where the embedding of the first token (\texttt{[CLS]} token) is used as the feature for \modelname{BERT} models.
All training and evaluation was done in five-fold cross validation setting.

Following related work \autocite{GruverSMWZU24, song-etal-2023-honeybee}, for \modelname{Llama} finetunes we employed a rank-size of 32 and $\alpha=64$, with a batch size of 8 for 5 epochs and a cosine-annealed learning rate of 0.0003, with no bias applied and on a  \texttt{CAUSAL\_LM} task. 
We accumulated gradients over 4 steps and employed gradient checkpointing. The learning rate was set to \num{3e-4} with a cosine scheduler, a warmup ratio of 0.03, and a weight decay of 0.001. Optimization was performed using the \texttt{paged\_adamw\_32bit} optimizer. A maximum gradient norm of 0.3 was maintained to ensure stable training.

Due to computational limitations, a comprehensive hyperparameter optimization was infeasible. However, for the \modelname{Llama} fine-tuning, some experimentation was performed with the \gls{lora} rank, learning rate, and warmup ratio. We found no significant performance difference for rank and learning rate (with results falling within the 5-fold cross-validation error bars).  The warmup ratio, however, was more sensitive, as smaller warmup steps resulted in more invalid generations. The remaining parameters were set following the cited work\autocite{GruverSMWZU24, song-etal-2023-honeybee} and the authors observed no trend that suggest that different parameters would alter the effects seen in the experiments.

\section*{Acknowledgments}

The research of N.A. and K.M.J. was supported by the Carl-Zeiss Foundation as well as Intel and Merck via the AWASES research center.  K.M.J.\ is part of the NFDI consortium FAIRmat funded by the Deutsche Forschungsgemeinschaft (DFG, German Research Foundation) – project 460197019. This work was also supported by the Helmholtz Association’s Initiative and Networking Fund on the HAICORE@FZJ partition. In addition, we thank André Sternbeck for support with computational infrastructure and acknowledge use of the \enquote{Draco} cluster at the Friedrich Schiller university of Jena as well as compute ressources at TU Ilmenau.
The authors thank Reza Aliakbari for his contributions to the implementation of material text representations.

We also thank Adrian Mirza, Sreekanth Kunchapu,
Mara Schilling-Wilhelmi, Martiño Rios Garcia, Anagha Aneesh, Meiling Sun,  Gordan Prastalo and Sadra Aghajani for their feedback on the draft.

\section*{Data availability}
To facilitate the benchmarking and reproducibility of our work, we have provided the datasets used in this work on HuggingFace.\autocite{nawaf_alampara_2024}

\section*{Code availability}

The code for using MatText is released under MIT license with tutorials and documentation at \url{https://github.com/lamalab-org/MatText}.  

\clearpage
\printbibliography[title=References]
\end{refsection}

\clearpage

\begin{refsection}

\section{Appendix}

\subsection{Related works on modeling materials using language models}
\label{subsec:material_modelling}

Language modeling has emerged as a promising method for predicting protein structure and function, representing amino acid sequences as text \autocite{Ruffolo_2024, Rives_2021, Lin_2023, Elnaggar_2022, xu2023protst}. 
While some research suggests language models capture structural information from sequence data \autocite{vig2021bertology}, others find their performance on many downstream tasks does not consistently scale with pretraining \autocite{Li_2024}. 
Similarly, text-based representations like SMILES \autocite{weininger1988smiles} and SELFIES \autocite{krenn2020self, cheng2023group} have been developed for molecules \autocite{Krenn_2022, bran2023transformers, Cadeddu_2014, Frey_2023, White_2023, noutahi2023gotta}, enabling language modeling for tasks such as synthesis planning \autocite{Pesciullesi_2020, Schwaller_2019}, property prediction \autocite{chithrananda2020chemberta, Wang_2019, ahmad2022chemberta2, balaji2023gptmolberta}, and conditional molecule generation \autocite{Born_2023, Bagal_2021, Grisoni_2023, ghugare2023searching}.

While protein and molecular text representations offer inspiration, materials science presents unique challenges due to properties depending on 3D structure and periodic repetition \autocite{Hoffmann_1987}, thereby complicating the development textual representations. 
Nevertheless, various approaches have been proposed, such as Robocrystallographer for human-readable crystal descriptions\autocite{ganose2019robocrystallographer}, used in predictive models \autocite{Qu_2024, Sayeed_2023, korolev2023toward}, and specialized representations like MOFid for specific material classes\autocite{Bucior_2019, Cao_2023}. 
However, no comprehensive universal representation has emerged for materials, making language modeling in this field significantly more challenging.

\subsection{Inductive Biases for Material Modeling} 
\label{appendix_sec:inductive_biases}
The modeling of physical systems can often benefit from the inclusion of physical background knowledge as inductive bias. 
Locality, smoothness, and symmetry are the most widely used inductive biases \autocite{Musil_2021}. 
Locality is commonly incorporated using a distance cutoff and rationalized with the nearsightedness principle of quantum mechanics \autocite{prodan2005nearsightedness}.
Related to this is using coarse-grained molecular motifs as inductive bias \autocite{sommer2023power, Cheng_2023}. 
Symmetry has been incorporated in many of the most performant models by designing invariant or equivariant features \autocite{Langer_2022, Musil_2021} or model architectures \autocite{batatia2022mace, satorras2021n, Batzner_2022, thomas2018tensor, duval2023hitchhiker}. 
Previous work has indicated that for certain phenomena (e.g., when all structures in a dataset are in the ground state), composition might implicitly encode geometric information \autocite{tian2022information, Jha_2018, Wang_2021}. 

\subsection{Representations}
\label{appendix_sec:representation}

\mattext encompasses nine distinct text-based representations for material systems, including several novel representations. 
Each representation incorporates unique inductive biases that capture relevant information and integrate prior physical knowledge about materials. \Cref{tab:inductive_biases} summarizes the inductive biases in each representation.

\begin{table*}[ht]
\label{tab:inductive_biases}
\centering
\caption{\textbf{\mattext Representations considered in this work and the inductive biases they encode.} Representations are classified into three broader categories based on the information it encodes.}
\label{tab:representations}
\begin{adjustbox}{max width=\linewidth}
\begin{tabular}{lcccccc}
\toprule
\textbf{Representation} & \textbf{Stoichiometry} & \textbf{Bonding} & \textbf{Geometry} & \textbf{Symmetry} & \textbf{Periodicity} & \textbf{Coarse‑Graining} \\
\midrule
\multicolumn{7}{c}{\textbf{Composition Representations}} \\
\midrule
\comp & \cmark & & & & & \\
\atoms & \cmark & & & & & \\
\atomsparams & \cmark & & & & \cmark & \\
\midrule
\multicolumn{7}{c}{\textbf{Local‑Geometry Representations}} \\
\midrule
\slices & \cmark & \cmark & & & \cmark & \\
\localenv & \cmark & \cmark & & \cmark & & \cmark \\
\midrule
\multicolumn{7}{c}{\textbf{Geometry Representations}} \\
\midrule
\crystaltext & \cmark & & \cmark & & \cmark & \\
\cifpone & \cmark & \cmark & \cmark & & \cmark & \\
\cifsym & \cmark & & \cmark & \cmark & \cmark & \\
\bottomrule
\end{tabular}
\end{adjustbox}
\end{table*}

\paragraph{Composition} represents the most basic level of material description, providing only the stoichiometric formula that indicates which elements are present and their relative ratios.
Prior work has shown that in certain cases, material composition alone can be predictive for various materials properties \autocite{tian2022information}. 
Hence, we also consider the composition in customary Hill notation \autocite{Hill_1900}.

\begin{practicalbox}[Example: \comp]

\begin{itemize}
\item Li7La3Zr2O12

\end{itemize}
\end{practicalbox}

\paragraph{Atom Sequence}
 To investigate the effect of the representation of compositional information, we explicitly list all the atoms present within the unit cell to eliminate any confusion that might arise from interpreting numbers as stoichiometric coefficients. 
 Concretely, structures are represented by listing each atom symbol $n$ times to denote repetition within the unit cell structure. 
 This representation is an intermediate representation between composition and representation containing information related to bonding or periodicity.

\begin{practicalbox}[Example: \atoms]

\begin{itemize}
\item Li Li Li Li Li Li Li La La La Zr Zr O O O O O O O O O O O O

\end{itemize}
\end{practicalbox}

\paragraph{Atom Sequence++}
We incorporate lattice parameters sequentially into the Atom Sequence to ablate the effect of having unit cell dimensions. 
This representation bridges the gap between purely compositional and fully structural descriptions. 
Although the position of atoms is not available, it contains information related to periodicity.

\begin{practicalbox}[Example: \atomsparams]

\begin{itemize}
\item Li Li Li Li Li Li Li La La La Zr Zr O O O O O O O O O O O O 11.3 11.3 11.3 108.3 108.3 111.6

\end{itemize}
\end{practicalbox}

\paragraph{SLICES}
In addition to composition information, SLICES encompasses the composition and bonding of atoms within and across the unit cell. 
It is an invariant and invertible string representation without explicit information about the atom coordinates  \autocite{xiao2023invertible}. 
The representation is a single-line string starting with elemental symbols within the unit cell and followed by bond descriptions in the format \texttt{uvxyz}. 
Here, \texttt{u} and \texttt{v} represent node indices, while \texttt{xyz} denotes the direction of the unit cell necessary to establish each bond connection across the unit cell boundaries.

\begin{practicalbox}[Example: \slices]

\begin{itemize}
\item Mg Ta Pt 0 2 - - o 0 2 - o - 0 2 - o o 0 2 o - - 0 2 o - o 0 2 o o - 0 1 - - o 0 1 - o - 0 1 o - - 1 2 o o o 

\end{itemize}
\end{practicalbox}

\paragraph{Local-Env}
We also report a new text representation inspired by the frequently used inductive bias of locality and Pauling's rule of parsimony, which states that local environments tend to be redundant \autocite{Pauling1960-it}. 
To derive the local environments, we perform the coordination environment analysis reported by \textcite{Waroquiers_2020}, derive Wyckoff labels using \spglib \autocite{togo2018texttt}, and SMILES using \openbabel \autocite{O_Boyle_2011}.
We prefix the representation using the spacegroup symbol and then list the Wyckoff label and SMILES separated by line breaks for each local environment. 
This representation features inductive biases related to local geometry. 

\begin{practicalbox}[Example: \localenv]

\begin{itemize}
\item R3m\\Ta (1a) [Ta]\#[Pt]\\Pt (1a) [Ta]\#[Pt]\\Mg (1a) [Ta][Mg][Ta].[Ta].[Pt].[Pt].[Pt]

\end{itemize}
\end{practicalbox}

\paragraph{Crystal-Text-LLM} This representation is a condensed version of the CIF, which includes only the parameters necessary for building the crystal structure\autocite{GruverSMWZU24} (without additional syntax of the CIF). 
Given the lattice parameters of the unit cell, atom types, and their coordinates, the bulk material structure can be represented as a listing of element symbols and coordinates separated by linebreaks that are prefixed by the list of lattice parameters (cell lengths and angles). 
\crystaltext is part of the representation group incorporating geometry inductive biases (spatial information) but they also have composition information (categorical information).

\begin{practicalbox}[Example: \crystaltext]

\begin{itemize}
\item 3.5 4.2 4.4\\90 90 90\\Ta\\0.76 0.12 0.00\\Ta\\0.00 0.12 0.18\\V\\0.00 0.00 0.00\\Ga\\0.76 0.00 0.18

\end{itemize}
\end{practicalbox}

\paragraph{CIF P$_1$} Crystallographic Information Files (CIFs) are a standard way to archive structural data in crystallography \autocite{hall1991crystallographic}. CIF P$_1$ represents the full Crystallographic Information File format using the primitive space group P$_1$.
They have been previously used for generating materials by fine-tuning \gls{llm}  \autocite{flam2023language, ding2025matexpert} or pretraining small GPT models \autocite{antunes2023crystal}. 
In the CIF P$_1$ representation, the crystal structure is represented in the lowest symmetry (P$_1$ space group).  
This means that if there is any symmetry in the crystal structure, it is not explicitly defined.  \cifpone is part of the representation group incorporating geometry inductive biases (spatial information) but they also have composition information (categorical information). 
Contrasting the \cifpone and the \crystaltext representations allows us to obtain insights into the importance of the compactness of representations.

\begin{practicalbox}[Example: \cifpone]

\begin{itemize}
\item data\_MgTaPt\\\_symmetry\_space\_group\_name\_H-M   'P 1'\\\_cell\_length\_a   4.32\\\_cell\_length\_b   4.32\\\_cell\_length\_c   4.32\\\_cell\_angle\_alpha   60.0\\\_cell\_angle\_beta   60.0\\\_cell\_angle\_gamma   60.0\\\_symmetry\_Int\_Tables\_number   1\\\_chemical\_formula\_structural   MgTaPt\\\_chemical\_formula\_sum   'Mg1 Ta1 Pt1'\\\_cell\_volume   56.85\\\_cell\_formula\_units\_Z   1\\loop\_\\ \_symmetry\_equiv\_pos\_site\_id\\ \_symmetry\_equiv\_pos\_as\_xyz\\  1  'x, y, z'\\loop\_\\ \_atom\_site\_type\_symbol\\ \_atom\_site\_label\\ \_atom\_site\_symmetry\_multiplicity\\ \_atom\_site\_fract\_x\\ \_atom\_site\_fract\_y\\ \_atom\_site\_fract\_z\\ \_atom\_site\_occupancy\\  Mg  Mg0  1  0.0  0.0  0.0  1.0\\  Ta  Ta2  1  0.58  0.58  0.58  1.0\\  Pt  Pt1  1  0.53  0.53  0.53  1.0

\end{itemize}
\end{practicalbox}

\paragraph{CIF symmetrized} This CIF representation represents the asymmetric unit and list the symmetry operations that can be applied to fill the unit cell by generating all equivalent positions.
It typically contains fewer lines describing atoms' positions than the CIF in P$_1$ but extra text describing the symmetry operations. 
Thus, this representation can contain more tokens for some structures than the P$_1$ variant (CIF P$_1$). 
This representation allows us to elucidate the importance of explicit symmetry information alongside positional information. 
\cifsym is part of the representation group incorporating geometry inductive biases (spatial information) but they also have composition information (categorical information). 

\begin{practicalbox}[Example: \cifsym]

\begin{itemize}
\item data\_MgTaPt\\\_symmetry\_space\_group\_name\_H-M   R3m\\\_cell\_length\_a   4.32\\\_cell\_length\_b   4.32\\\_cell\_length\_c   10.57\\\_cell\_angle\_alpha   90.0\\\_cell\_angle\_beta   90.0\\\_cell\_angle\_gamma   120.0\\\_symmetry\_Int\_Tables\_number   160\\\_chemical\_formula\_structural   MgTaPt\\\_chemical\_formula\_sum   'Mg3 Ta3 Pt3'\\\_cell\_volume   170.55\\\_cell\_formula\_units\_Z   3\\loop\_\\ \_symmetry\_equiv\_pos\_site\_id\\ \_symmetry\_equiv\_pos\_as\_xyz\\  1  'x, y, z'\\  2  '-y, x-y, z'\\  3  '-x+y, -x, z'\\  4  '-y, -x, z'\\  5  '-x+y, y, z'\\  6  'x, x-y, z'\\  7  'x+1/3, y+2/3, z+2/3'\\  8  '-y+1/3, x-y+2/3, z+2/3'\\  9  '-x+y+1/3, -x+2/3, z+2/3'\\  10  '-y+1/3, -x+2/3, z+2/3'\\  11  '-x+y+1/3, y+2/3, z+2/3'\\  12  'x+1/3, x-y+2/3, z+2/3'\\  13  'x+2/3, y+1/3, z+1/3'\\  14  '-y+2/3, x-y+1/3, z+1/3'\\  15  '-x+y+2/3, -x+1/3, z+1/3'\\  16  '-y+2/3, -x+1/3, z+1/3'\\  17  '-x+y+2/3, y+1/3, z+1/3'\\  18  'x+2/3, x-y+1/3, z+1/3'\\loop\_\\ \_atom\_site\_type\_symbol\\ \_atom\_site\_label\\ \_atom\_site\_symmetry\_multiplicity\\ \_atom\_site\_fract\_x\\ \_atom\_site\_fract\_y\\ \_atom\_site\_fract\_z\\ \_atom\_site\_occupancy\\  Mg  Mg0  3  0.0  0.0  0.0  1.0\\  Ta  Ta1  3  0.0  0.0  0.42  1.0\\  Pt  Pt2  3  0.0  0.0  0.47  1.0

\end{itemize}
\end{practicalbox}

\paragraph{Z-matrix}
The z-matrix is a representation widely used as input for quantum mechanical simulations of small molecules (but not materials).\autocite{cramer2013essentials} 
It leverages internal coordinates and is hence invariant with respect to translation or rotation. 
The internal coordinates used in a z-matrix are bond distances, angles, as well as dihedral angles. 
As all of these internal coordinates are defined with respect to neighboring atoms, the representation implicitly also encodes bonds. 
Here, we define the z-matrix based on the atoms within one unit cell. \zmatrix is part of the representation group incorporating geometry inductive biases (spatial information) but they also have composition information (categorical information).

\begin{practicalbox}[Example: \zmatrix]

\begin{itemize}
\item Mg\\Ta 1 6.1\\Pt 2 0.5 1 0

\end{itemize}
\end{practicalbox}

Each of these representations, have different range of lengths. 
As the information content increase the representation length and the tokens required to model these system increases. 
Since there is a big variation in the representation length, we have modeled them with different context lengths, but fixed amongst representations. 
Appendix \cref{table:context_lengths} indicates the context length used for modeling different representations.
\begin{table}[h]
\centering
\caption{Representations and their corresponding context lengths.}
\begin{tabular}{cc}
\toprule
\textbf{Representation} & \textbf{Context Length} \\
\midrule
SLICES & 512 \\
Composition & 32 \\
Crystal-text-LLM & 512 \\
Z-Matrix & 512 \\
CIF P$_1$ & 1024 \\
CIF Symmetrized & 1024 \\
Atom Sequence & 32 \\
Atom Sequence++ & 32 \\
Local-Env & 512 \\
\bottomrule
\end{tabular}
\label{table:context_lengths}
\end{table}

\subsection{Attention Analysis}
\label{appendix_sec:attention_analysis}

The amount of attention received by different tokens can be interpreted as a measure of the relevance of different tokens contained in the representations (\cref{appendix_sec:representation}). 

The models here attend most to atomic symbols. This observation aligns with our primary findings, which show a strong correlation between leveraging compositional information (\cref{fig:representation-depedence}). 

Consistently, we also observe that numbers generally receive less attention. 
Overall, this supports the hypothesis that current models do not effectively utilize numerical information for learning complex geometric features.  

\begin{figure}[tbh]
    \centering
    \includegraphics[width=1\linewidth]{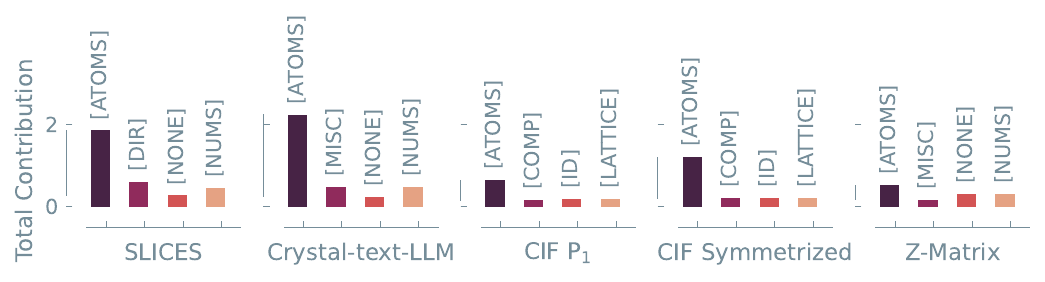}
    \caption{\textbf{Attention received by different types of tokens in different representations summed across all the heads and layers}. Composition related (categorical information) tokens receives highest contributions and surprisingly numbers receive less attention leading to spatial information not being leveraged in prediction}
    \label{fig:perovskite_attention_head}
\end{figure}

\paragraph{Token attention contribution calculation}
To perform this analysis, we first compute the contribution per token.

The element-wise multiplication of the attention matrix \( A^{(l,h)} \) and mask \( M_k \) gives the contribution of the attention scores for the token type \( k \):
\[
C^{(l,h)}_k = A^{(l,h)} \odot M_k
\]
Here, \( \odot \) denotes element-wise multiplication.

In this context, \( A^{(l,h)} \) represents the attention matrix for layer \( l \) and head \( h \), and \( M_k \) is the mask for token type \( k \) in tokenized material text representations. 
Tokens can be classified into different types for analytical purposes. 
For example, the SLICES representations can have tokens of the type \texttt{ATOMS}, \texttt{NUMS}, and \texttt{DIR}. 
Specifically, all atoms are classified under the \texttt{ATOMS} token type, numbers are classified under \texttt{NUM}, and \texttt{DIR} represents tokens defining the direction of bonds.

The mask \( M_k \) is defined as:
\[
M_k \in \{0,1\}^{T \times T},
\]
where \( M_k \) is a binary matrix taking values 0 or 1.

The dimension of \( M_k \) matches that of the attention weight matrix \( A^{(l,h)} \). Given that samples in the dataset may contain varying numbers of atoms, each sample can have different corresponding masks. 
To facilitate this analysis, the MatText tokenizers provide the functionality to generate a list of token types alongside the list of tokens. These token types are used dynamically to design masks for attention analysis.

\paragraph{Token Weight}

The percentage weight for token type \( k \) in layer \( l \) and head \( h \) is then calculated as:
\[
W^{(l,h)}_k = \frac{\sum_{i,j} \left( C^{(l,h)}_k(i,j) \right)}{\sum_{i,j} M_k(i,j)}.
\]

Here \(W^{(l,h)}_k \) is the percentage attention recieved by a particular token type \( k \) in layer \( l \) and head \( h \) during prediction and \( \sum_{i,j} \) denotes summing over all elements in the matrix.

\paragraph{Aggregation Across Folds}
Aggregates the attention weights for all the samples across multiple folds. T
his involves summing the weights for each token across all samples ($N$) and folds ($f$).
The total contribution of token type \( k \) across all folds is given by:
\[
T_k = \sum_{N} \sum_{f} \sum_{l,h} W^{(l,h,f)}_k.
\]

\paragraph{Results}
In the attention heat maps, we observe certain heads specialized to learn features from compositions, which is not the case for numbers, where we observe rather a dispersed nature in heads (\Cref{fig:perovskite_attention_head}). 
Previously, with unsupervised language modeling of proteins, the formation of such heads was associated with parts of the architecture concentrating on learning certain features \autocite{vig2021bertology}.
We observe that groups dedicated to learning numerical features do not emerge with pretraining.

\subsection{Tokenizer}
\label{appendix_sec:tokenizer}

Low performance of \glspl{llm} has often been associated to tokenizers where numbers might not be processed correctly due to the tokenization method \autocite{singh2024tokenization, zausinger2024regress}. For instance, in many default tokenization methods (e.g., BPE, single-digit tokenization), numbers are represented with varying numbers of tokens, which might make it more difficult for models to use them effectively.  
To address this issue, we implemented the tokenizer proposed by \textcite{Born_2023}, which preserves decimal order and also encodes the order of magnitude. 
We find that this change in tokenizer does not provide consistent improvements in modeling performance.

\begin{figure}[H]
    \centering
    \includegraphics[width=1\linewidth]{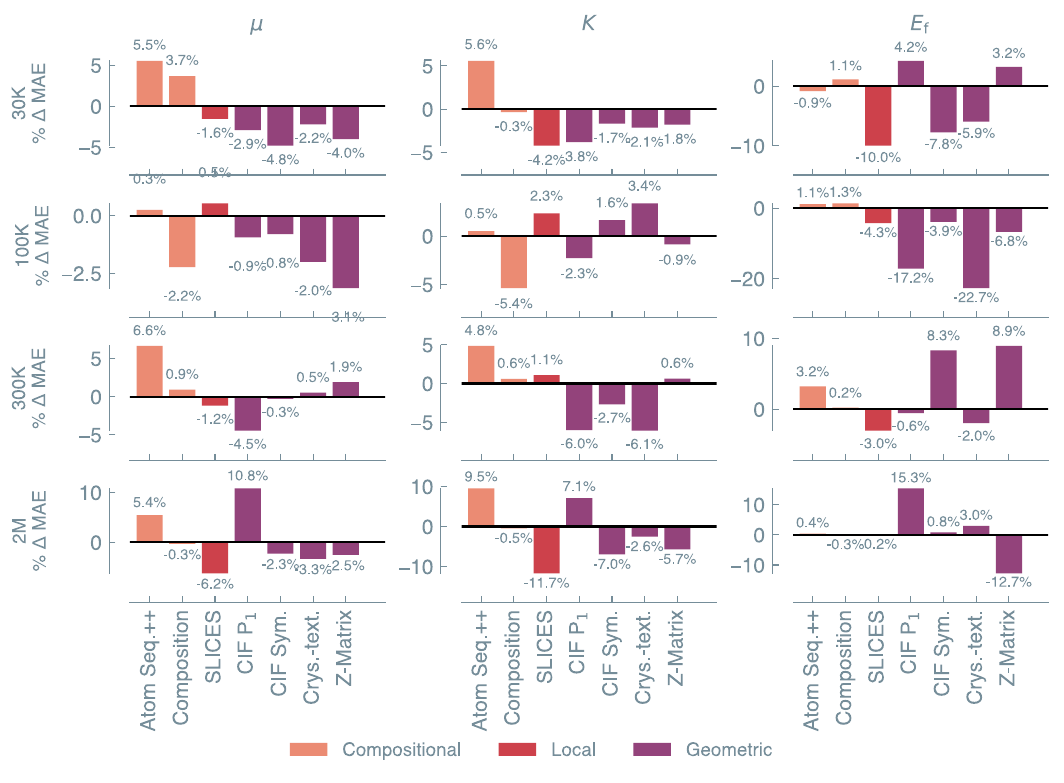}
    \caption{\textbf{Performance comparison between the RT tokenizer \autocite{Born_2023} and the regular tokenizer across different data scales and representation types.} Each subplot shows the percentage improvement in \gls{rmse} when using RT tokenizer compared to normal tokenizer, where positive values indicate RT tokenizer performs better. The plot is organized with four rows representing different training dataset sizes (30K, 100K, 300K, 2M) and three columns representing different material properties: shear modulus ($\mu$), bulk modulus ($\kappa$), and formation energy of perovskites ($E_{f}$). 
    Bars are colored by representation group: orange for compositional representations (\atomsparams, \comp), red for local representations (\slices), and purple for geometric representations (\crystaltext, CIF variants, \zmatrix).
    }
    \label{fig:tokenizer_performance}
\end{figure}

The RT tokenizer shows predominantly mixed effects on model performance with high variability across properties and representations. 
While there are minor improvements for some representations at large scales (particularly for geometric representations like \zmatrix), the overall pattern reveals inconsistent and context-dependent performance changes. 
\gls{fe}  exhibits particularly high variation, with some representations showing substantial improvements while others demonstrate significant degradations. 
\gls{bm} shows the most consistent behavior but still with mixed results, while \gls{sm} demonstrates largely neutral to modest effects. 
The amplification of variability at larger scales (300K, 2M) suggests that RT tokenizer's impact becomes more unpredictable rather than universally beneficial with increased training data.

\subsection{Coordinate-Category Cliff in material property prediction}
\label{sec:ssa-property}

\begin{figure}[H]
    \centering
    \includegraphics[width=1\linewidth]{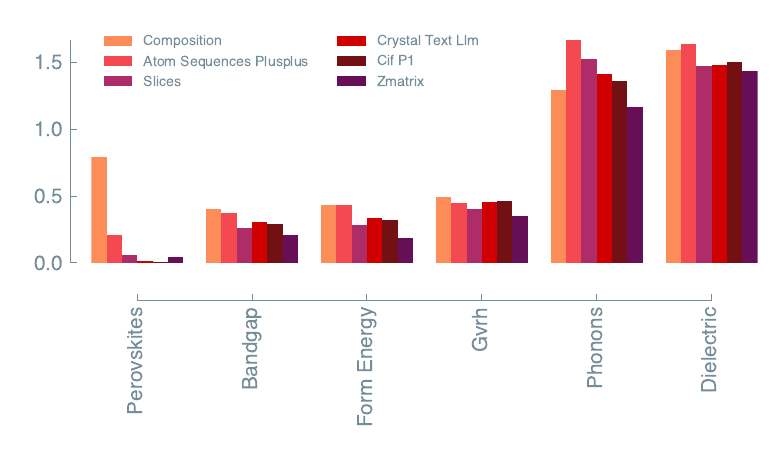}
    \caption{\textbf{\gls{ccc} for different representations for different properties.} Each property is a dataset with multiple crystal structures. All representations are created and modeled using \mattext framework}. Dataset with complex structures like dielectric materials have higher \gls{ccc} values, while simpler structures like that of small perovskite strucutre dataset show lower \gls{ccc}.
    \label{fig:ssa-materials}
\end{figure}

We compute \gls{ccc} for different  representations with different inductive biases for different material science property dataset. Again for the purpose of quantifying this metric, we can compute \gls{ccc} as \gls{coc} $-$ \gls{cac}. Positive and big \gls{ccc} indicate that the model finds it substantially harder to learn from coordinate information than from category information. In \Cref{fig:ssa-materials} we have plotted \gls{ccc} for material property prediction.
Irrespective of the representations we find that for all datasets, models struggle leveraging coordinate information from the representations.

\subsection{Comparison of \gls{ccc} with \modelname{n-gram} models}

\begin{figure}[H]
    \centering
    \includegraphics[width=1\linewidth]{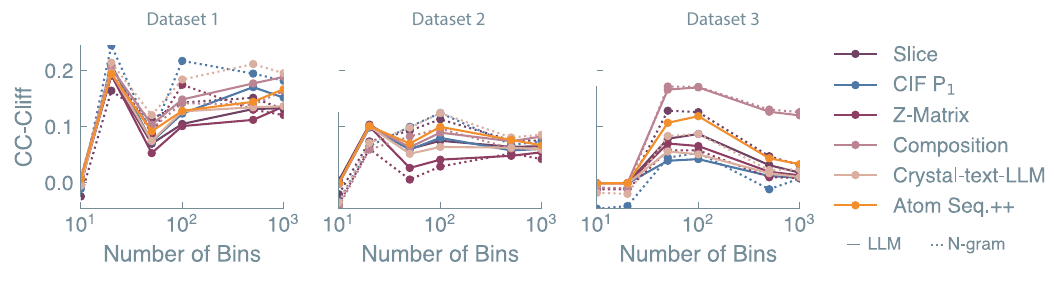}
    \caption{CC-Cliff in hypothetical potential energy prediction as a function of
binning the potential across different datasets and representations. The figure illustrates the CC-Cliff for models trained on different representations tasked with predicting hypothetical potential energy. The analysis is performed across three distinct datasets. The plot shows almost identical behavior of language models to that
of \modelname{n-gram} models suggesting that \gls{llm} inherit the properties of \modelname{n-gram} for tasks involving
coordinate and category data irrespective of representation}
    \label{fig:ngram_allrep}
\end{figure}

In \cref{fig:ngram_allrep} we find that for all datasets we analyzed, the behavior of \modelname{n-gram} models closely follows that of the \glspl{llm} trained on different representations. More details on the methodology is \cref{subsubsec:ngram_rationale}. 

\subsection{Relationship of \gls{ccc} with diversity}

\begin{figure}[H]
    \centering
    \begin{subfigure}[b]{0.9\textwidth} 
        \centering
        \includegraphics[width=\textwidth]{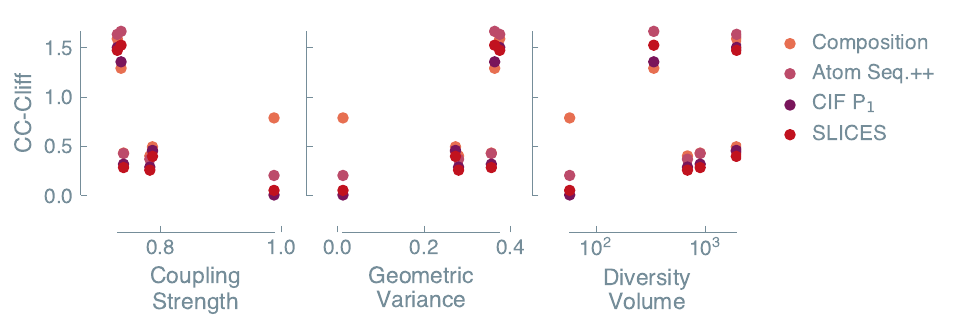}
        \caption{\textbf{Comparison between structural diversity and the \gls{ccc} across different representations.} 
        Each point represents the \gls{ccc} of the representations.  Higher coupling strength indicates that the composition more strictly constrains possible structural arrangements. Datasets used in this analysis are, Perovskites, MP-Formation energy, MP-Bandgap, GVRH, phonons and shear modulus. 
        Error bars show the standard deviation within each representation group. }
        \label{fig:ssa_diversity}
    \end{subfigure}
    \par\vspace{1cm} 

    \begin{subfigure}[b]{0.9\textwidth} 
        \centering
        \includegraphics[width=\textwidth]{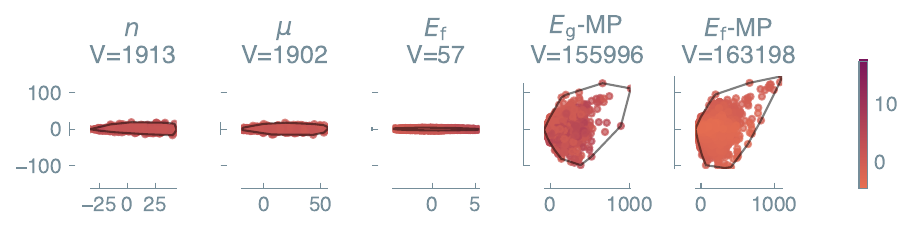}
        \caption{\textbf{Structural variations across different material property dataset.} Diversity volume is quantified using the convex hull volume of the representation space. Higher volumes indicate greater structural diversity in the dataset, revealing how different representation approaches handle increasingly complex structural variations.} 
        \label{fig:structural_variations_all_property}
    \end{subfigure}
    \caption{\gls{ccc} shown by representations against different diversity metrics for different material datasets and method employed to compute diversity for datasets}
    \label{fig:combined_figures}
\end{figure}

From Appendix \cref{{sec:ssa-property}} we see that \gls{ccc}   exhibits correlation with different datasets, however it is not clear why certain datasets show bigger \gls{ccc}, i.e., why in certain crystal structure datasets it seems difficult for \gls{llm} to learn the geometric relationship more. 
This could be because of the type of crystal structures, often some materials has complex coordination and hence can have complex geometric relationships. 
To understand the impact of the diversity of the structures in dataset, we benchmarked the diversity of the datasets by computing the coupling strength, geometric variance and diversity volume.
The diversity volume in different datasets is quantified using convex hull volume calculated from the first two components of PCA projection of material embeddings. We apply the embeddings obtained using the \modelname{ChgNet} \autocite{deng2023chgnet} checkpoint v0.3. 
The volume (or area in 2D) of the convex hull provides a measure of how \enquote{spread out} the structures are in the latent space. 
A larger volume indicates more structural diversity. \Cref{fig:structural_variations_all_property} shows convex hull for embedding space using \modelname{ChgNet} model.

Geometric variance how much the geometric properties vary among structures with the same composition. For structures with identical composition,\\ 
let $\mathbf{G}_i = [g_{i,1}, g_{i,2}, \ldots, g_{i,7}]$ be the geometric fingerprint of structure $i$, where each $g_{i,j}$ represents a statistical measure of pairwise atomic distances (min, max, mean, median, std, Q25, Q75). The geometric variance is:

\begin{equation}
\text{geometric variance}_c = \frac{1}{7} \sum_{j=1}^{7} \text{Var}(g_{i,j})
\end{equation}

Lower geometric variance indicates that composition more strongly determines structure, while higher variance suggests greater structural diversity for the same composition. Coupling strength (\cref{eq:coupling_strength}) measures how strongly composition determines structure.
\begin{equation}
    \label{eq:coupling_strength}
   \text{coupling strength} = \frac{1}{1 + \text{mean\_geometric\_variance}} 
\end{equation}

We computed the diversity metrics and compared \gls{ccc} for each of these metrics for different representations.  
We see from \cref{{fig:ssa_diversity}} that \gls{ccc} in general increases with diversity but not consistently.

\clearpage
\printbibliography[title=Appendix References]
\end{refsection}

\footnotesize
\normalsize

\clearpage
\printnoidxglossary[type=\acronymtype]
\printnoidxglossary[sort=letter]

\end{document}

%% file: preamble.tex
\usepackage[english]{babel}
\usepackage{geometry}  
\usepackage{graphicx}      
\usepackage{tabularx}      
\usepackage{amssymb}
\usepackage{amsmath}
\usepackage{eulervm}
\usepackage{xltabular}
\usepackage[hyphens]{url}
\usepackage[hidelinks]{hyperref}
\hypersetup{breaklinks=true}
\usepackage[backend=biber,style=nature, doi=false,isbn=false, date=year, url=false, autopunct=true]{biblatex}
\makeatletter
\renewbibmacro*{textcite}{%
  \iffieldequals{namehash}{\cbx@lasthash}
    {\usebibmacro{cite:comp}}
    {\usebibmacro{cite:dump}%
     \ifbool{cbx:parens}
       {\bibclosebracket\global\boolfalse{cbx:parens}}
       {}%
     \iffirstcitekey
       {}
       {\textcitedelim}%
     \usebibmacro{cite:init}%
     \ifnameundef{labelname}
       {\printfield[citetitle]{labeltitle}}
       {\printnames{labelname}}%
     \addspace
     \ifnumequal{\value{citecount}}{1}
       {\usebibmacro{prenote}}
       {}%
     \mkbibsuperscript{\usebibmacro{cite:comp}}%
     \stepcounter{textcitecount}%
     \savefield{namehash}{\cbx@lasthash}}}
\makeatother

\usepackage{csquotes}
\usepackage{cleveref}
\usepackage{siunitx}
\usepackage{geometry}
\usepackage{multirow}
\usepackage[dvipsnames,table,xcdraw]{xcolor}
\definecolor{tumblue}{rgb}{0.5450980392,0.4901960784,0.6862745098}
\definecolor{mattext_red}{rgb}{0.4745098039215686,0.08235294117647059,0.3568627450980392}

\usepackage{xspace}
\usepackage{caption}
\usepackage{subcaption}
\usepackage{authblk}

\usepackage{marvosym}

\usepackage{sectsty} 

\sectionfont{\color{mattext_red}\selectfont\sffamily}
\subsectionfont{\color{mattext_red}\selectfont\sffamily}
\subsubsectionfont{\color{mattext_red}\selectfont\sffamily}
\paragraphfont{\selectfont\sffamily}
\subparagraphfont{\color{gray}\selectfont\sffamily}

\definecolor{halfgray}{gray}{0.55}

\IfFileExists{./fonts/Heliotrope-Bold.otf}{
    \usepackage{fontspec}
    \setmainfont{Heliotrope}[Extension = .otf, UprightFont = *-Regular, ItalicFont = *-Italic, BoldFont=*-Bold, Path = ./fonts/]
}{
    \usepackage{biolinum}
    
}

\usepackage[labelfont=bf]{caption}
\usepackage[nonumberlist,acronym]{glossaries}
\input{acronyms}
\makenoidxglossaries 
\setacronymstyle{long-short}

 \usepackage{microtype}
 \usepackage{fontawesome}

\definecolor{linkcolor}{RGB}{0, 102, 204}

\newcommand{\comp}{Composition\xspace}
\newcommand{\atoms}{Atom Sequence\xspace}
\newcommand{\atomsparams}{Atom Sequence++\xspace}
\newcommand{\slices}{SLICES\xspace}
\newcommand{\localenv}{Local-Env\xspace}
\newcommand{\cifsym}{Cif Symmetrized\xspace}
\newcommand{\cifpone}{Cif P$_1$\xspace}
\newcommand{\zmatrix}{Z-Matrix\xspace}
\newcommand{\crystaltext}{Crystal-Text-LLM\xspace}

\newcommand{\comprepgroup}{Composiiton, Atom Sequence, Atom Sequence++}
\newcommand{\localrepgroup}{SLICES, Local-Env}
\newcommand{\geomrepgroup}{Cif P$_1$,Cif Symmetrized, Z-Matrix, Crystal-Text-LLM}

\newcommand{\spglib}{\texttt{spglib}\xspace}
\newcommand{\openbabel}{\texttt{openbabel}\xspace}

\newcommand{\mattext}{MatText\xspace}

\definecolor{implicationbg}{RGB}{245, 245, 245}

\newcommand{\modelname}[1]{\texttt{#1}}

%% file: acronyms.tex
\newacronym{llm}{LLM}{large language model}
\newacronym{gnn}{GNN}{graph neural network}
\newacronym{ccc}{CC-Cliff}{Coordinate-Category Cliff}
\newacronym{ssa}{SSA}{spatial-semantic asymmetry}
\newacronym{ccc-l}{CC-Cliff}{Coordinate-Category Cliff in learning}
\newacronym{coc}{$L_{\mathrm{coord}, \alpha}$}{coordinate contribution}
\newacronym{cac}{$L_{\mathrm{cat}, \alpha}$}{category contribution}
\newacronym{sc}{$G_\alpha$}{spatial contribution}
\newacronym{cc}{$C_\alpha$}{semantic contribution}
\newacronym{gcmg}{GCMG}{geometry composition modeling gap}
\newacronym{ml}{ML}{machine learning}
\newacronym{agi}{AGI}{artificial general intelligence}
\newacronym{api}{API}{application programming interface}
\newacronym{mcq}{MCQ}{multiple-choice question}
\newacronym{rest}{REST}{representational state transfer}
\newacronym{orm}{ORM}{object relational mapping}
\newacronym{dai}{DAI}{daily allowed intake}
\newacronym{ghs}{GHS}{globally harmonized system of classification and labelling of chemicals}
\newacronym{who}{WHO}{World Health Organization}
\newacronym{nmr}{NMR}{nuclear magnetic resonance}
\newacronym{helm}{HELM}{Holistic Evaluation of Language Models}
\newacronym{smiles}{SMILES}{simplified molecular input line-entry system}
\newacronym{pca}{PCA}{principal component analysis}
\newacronym{iupac}{IUPAC}{International Union of Pure and Applied Chemistry}
\newacronym{json}{JSON}{JavaScript object notation}
\newacronym{rag}{RAG}{retrieval augmented generation}
\newacronym{ece}{ECE}{expected calibration error}
\newacronym{cot}{CoT}{chain of thought}
\newacronym{rlhf}{RLHF}{reinforcement learning with human feedback}
\newacronym{vllm}{VLLM}{Vision Large Language Model}
\newacronym{mof}{MOF}{metal-organic framework}
\newacronym{xrd}{XRD}{X-ray diffraction}
\newacronym{afm}{AFM}{atomic force microscopy}
\newacronym{ms}{MS}{mass spectrometry}
\newacronym{cif}{CIF}{Crytallographic Information Files}
\newacronym{qa}{QA}{Question-Answer Pairs}
\newacronym{fe}{$E_{f}$}{Formation Energy}
\newacronym{bm}{$\kappa$}{Bulk Modulus}
\newacronym{sm}{$\mu$}{Shear Modulus}
\newacronym{rmse}{RMSE}{Root Mean Square Error}
\newacronym{mae}{MAE}{Mean Absolute Error}
\newacronym{uff}{UFF}{Universal Force Field}
\newacronym{tfidf}{TF-IDF}{Term Frequency-Inverse Document Frequency}
\newacronym{qlora}{QLoRA}{quantized low-rank adaptation}
\newacronym{lora}{LoRA}{low-rank adaptation}

%% file: authors.tex
\author[1, \Letter]{Nawaf~Alampara~\orcidlink{0009-0001-7697-7315}}

\author[2]{Santiago Miret}

\author[1,3,4,5, \Letter]{Kevin~Maik~Jablonka~\orcidlink{0000-0003-4894-4660}}

\affil[1]{Laboratory of Organic and Macromolecular Chemistry (IOMC), Friedrich Schiller University Jena, Humboldtstrasse 10, 07743 Jena, Germany}

\affil[2]{Intel labs}

\affil[3]{Center for Energy and Environmental Chemistry Jena (CEEC Jena), Friedrich Schiller University Jena, Philosophenweg 7a, 07743 Jena, Germany}
\affil[4]{Helmholtz Institute for Polymers in Energy Applications Jena (HIPOLE Jena), Lessingstrasse 12-14, 07743 Jena, Germany}

\affil[5]{Jena Center for Soft Matter (JCSM), Friedrich Schiller University Jena, Philosophenweg 7, 07743 Jena, Germany}

\affil[\Letter]{\texttt{nawaf.alampara@uni-jena.de}, \texttt{mail@kjablonka.com}}